\documentclass[aps,nofootinbib,superscriptaddress,twocolumn,preprintnumbers,hyperref]{revtex4}
\usepackage{graphicx,amsmath,hyperref,amssymb,bbm}
\hypersetup{pdfstartview=FitH}
\newcommand{\Feyn}[1]{#1\kern-0.65em/}
\def\be{\begin{eqnarray}}
\def\ee{\end{eqnarray}}

\def\vol{{\mathtt{v}}}

\newcommand{\C}{\mathbb{C}}

\newcommand{\fref}[1]{Fig.\,\ref{#1}}

\begin{document}
\title{ \Large Cosmological constant in spinfoam cosmology}
    \author{Eugenio Bianchi}
    \email{bianchi@cpt.univ-mrs.fr}
\address{Centre de Physique Th\'eorique de Luminy\footnote{Unit\'e de recherche (UMR 6207) du CNRS et des Universit\'es de Provence (Aix-Marseille I), de la M\'editerran\'ee (Aix-Marseille II) et du Sud (Toulon-Var); affili\'e \`a la FRUMAM (FR 2291).\vskip2pt}
        , Case 907, F-13288 Marseille, EU;}
   
    \author{Thomas Krajewski}
    \email{krajewski@cpt.univ-mrs.fr}
\address{Centre de Physique Th\'eorique de Luminy\footnote{Unit\'e de recherche (UMR 6207) du CNRS et des Universit\'es de Provence (Aix-Marseille I), de la M\'editerran\'ee (Aix-Marseille II) et du Sud (Toulon-Var); affili\'e \`a la FRUMAM (FR 2291).\vskip2pt}
        , Case 907, F-13288 Marseille, EU;}
\address{Laboratoire de Physique Th\'eorique, Universit\'e Paris XI CNRS UMR 8627, 91405 Orsay Cedex, France;}

    \author{Carlo Rovelli}
    \email{rovelli@cpt.univ-mrs.fr}
\address{Centre de Physique Th\'eorique de Luminy\footnote{Unit\'e de recherche (UMR 6207) du CNRS et des Universit\'es de Provence (Aix-Marseille I), de la M\'editerran\'ee (Aix-Marseille II) et du Sud (Toulon-Var); affili\'e \`a la FRUMAM (FR 2291).\vskip2pt}
        , Case 907, F-13288 Marseille, EU;}
   
    \author{Francesca Vidotto}
    \email{vidotto@cpt.univ-mrs.fr}
\address{Centre de Physique Th\'eorique de Luminy\footnote{Unit\'e de recherche (UMR 6207) du CNRS et des Universit\'es de Provence (Aix-Marseille I), de la M\'editerran\'ee (Aix-Marseille II) et du Sud (Toulon-Var); affili\'e \`a la FRUMAM (FR 2291).\vskip2pt}
        , Case 907, F-13288 Marseille, EU;}
\address{Dipartimento di Fisica Nucleare e Teorica, Universit\`a degli Studi di Pavia,}
\address{ Istituto Nazionale di Fisica Nucleare, Sezione di Pavia, via A. Bassi 6, I-27100 Pavia, EU.}

\date{\today}

\begin{abstract}\noindent
We consider a simple modification of the amplitude defining the dynamics of loop quantum gravity, corresponding to the introduction of the cosmological constant, and possibly related to the $SL(2,\C)_q$  extension of the theory recently considered by Fairbairn-Meusburger and Han. We show that in the context of spinfoam cosmology, this modification yields the de Sitter cosmological solution. 
\noindent \end{abstract}
\maketitle

\section{Introduction}

The cosmological-constant term in the Einstein equations is an integral part of the theory that appears today to best describe Nature \cite{Lahav:ys,Bianchi:2010uw,Bianchi:2010kx}.  Incorporating this term into the covariant dynamics of loop quantum gravity (see \cite{Rovelli:2010vv,Rovelli:2010wq} and references therein) is important in order to get a realistic theory and also to control infrared divergences. But it is also important for formal reasons.  In particular, one of the elements of evidence that the covariant theory has the correct semiclassical limit, is its application to cosmology \cite{Bianchi:2010zs}.  So far this has been studied only in the absence of matter, but without matter and without cosmological constant, the only cosmological solution is flat space.  Recovering flat space is interesting, but is still weak evidence for the full classical limit.  

Here we consider a simple way of including the cosmological constant into the spinfoam vertex. We show that in cosmology this yields the de Sitter solution of the Einstein equations. 

The modification of the vertex that we consider is motivated by simple heuristic, based on the form of the cosmological constant term in the Hamiltonian constraint.  In two recent papers,  Fairbairn and Meusburger  \cite{Fairbairn:2010cp} and, independently, Han \cite{Han:2010pz}, have defined a modification of the spinfoam amplitude defining the dynamics of loop quantum gravity, by replacing $SL(2,\C)$ with its quantum deformation, developed in \cite{Noui:2002ag}.  It is reasonable to expect this modification to define quantum general relativity with a cosmological constant \cite{Turaev:1992hq,Perez:2010pm}.  Since the asymptotical analysis of the $q$-deformed vertex is not yet done, we cannot directly compare it with the the vertex we write here. Still, we argue below that if the $q$-deformed vertex behaves as expected in the large distance limit, the two definitions of the vertex should agree in this limit.

\section{Cosmological term in the full spinfoam amplitude}

Quantum-gravity transition amplitudes can be written in a number of equivalent forms \cite{Bianchi:2010mw,Rovelli:2010vv,Rovelli:2010wq}. Let us write the partition function in the spin-network basis as
\be
   Z_{\cal C}=\sum_{j_f,\vol_e} \prod_f (2j+1) \prod_v A_v(j_f,\vol_e). 
   \label{old}
\ee
Here $\cal C$ is a fixed two-complex, the sum is over a spin $j_f$ associated to each of its faces and an intertwiner $\vol_e$ associated to each of its edges $e$. The vertex amplitude $A_v(j_f,\vol_e)$ is a function of the spins and intertwiners adjacent to the vertex $v$.  It is convenient to choose a basis of intertwiners that diagonalizes the volume, and we indicate with $\vol_e$ the corresponding quantum number, which we take to be the eigenvalue (for simplicity of notation we disregard the eventual degeneracy).
We consider the following modification of the sum \eqref{old}, 
\be
   Z_{\cal C}=\sum_{j_f,\vol_e} \prod_f (2j+1)  \prod_e e^{i\lambda \vol_e} \prod_v A_v(j_f,\vol_e). 
\label{modificazione}
\ee

\vspace*{-2mm}\noindent 
where $\lambda$ is related to the cosmological constant. The relation between $\lambda$ and the standard cosmological constant $\Lambda$ is given below, together with a more precise justification for the form of the term added.

The heuristic for this definition is the following. In the canonical theory, the cosmological constant appears as an additive term to the gravitational Hamiltonian constraint, which multiples the 3-volume element. When deriving a path integral formulation of quantum theory \`a la Feynman by inserting resolutions of unity into the evolution operator, a potential term appears simply as a multiplicative exponential, because the potential is diagonal in the position basis. The cosmological constant term is diagonal in the spin-intertwiner basis.  It is therefore possible to insert the cosmological constant ``potential" as a multiplicative term along the spinfoam evolution, that is in between 4-cells, which is to say on 3-cells. The coupling is therefore very simple, and consists in weighting edge amplitudes with an exponential term which depends on the volume and the cosmological constant. 

Alternatively, the $q$-deformed version of the theory studied in  \cite{Fairbairn:2010cp,Han:2010pz} is expected to lead to a modification of the amplitude that corresponds to the addition of the cosmological term. In the large-$j$ regime, the conventional vertex amplitude converges to the Regge action of the 4-cell dual to the vertex, and the  $q$-deformed amplitude should converges to the Regge action of the 4-cell plus a cosmological term. This is given by the cosmological constant multiplying the four-volume of the 4-cell. But the boundary geometry is in the time gauge, where the Shift function vanishes and the Lapse function is equal to unity, and in this gauge the 4-volume is equal to the 3-volume.  The time gauge is also the common choice in the cosmological formalism. Hence we obtain again a modification of the vertex amplitude of the form \eqref{modificazione}.

\section{Cosmological model}

Let us apply the amplitude  \eqref{modificazione} to spinfoam cosmology, following 
\cite{Bianchi:2010zs}, to which we refer for the notation and the rest of the derivation.  By sandwiching the cosmological term $e^{i\lambda \vol_e}$ between coherent states, we obtain its contribution to the holomorphic form of the action. We are interested in the homogeneous and isotropic coherent states on the chosen graph, that here is a  ``dipole'' 
\!\!\!\!\begin{picture}(20,8)
	\put(08,3) {\circle*{3}} 
	\put(30,3) {\circle*{3}}  
	\qbezier(8,3)(19,18)(30,3)
	\qbezier(8,3)(19,-3)(30,3)
	\qbezier(8,3)(19,9)(30,3)
	\qbezier(8,3)(19,-11)(30,3)
\end{picture}~~~
.  For these states, the cosmological term is just a function of the spin
\be
     e^{i\lambda \vol_e}= e^{i\lambda \vol_o j^{3/2}}
\ee
where $\vol_o$ is the volume of a regular tetrahedron with faces having unit area. Therefore the effect of the cosmological constant is to modify the amplitude for a homogeneous isotropic state determined by the complex number $z$, where  $Re(z)\sim  \gamma  \dot a$, $Im(z)\sim a^2$, and $a$ is the scale factor,
\begin{eqnarray}
W(z) = \int_{SO(4)} dg \prod_{l=1,4}P_t(H_\ell,G) 
\end{eqnarray}
by replacing  the expression of $P_t(H_l,g)$ given by the 
equation after eq.(32) in \cite{Bianchi:2010zs} by 
\begin{eqnarray}
P_t(H_\ell,G) &=& 
\sum_{j} { (2j+1)} \, e^{-2t\hbar j(j+1)- i zj-i\lambda \vol_o j^{\frac32}}  \nonumber \\ && \hspace*{1em}\times \ \ 
{\rm Tr}\!
\left[
P_\ell Y^\dagger  D^{\scriptscriptstyle(j^{\!+}\!\!,j^{\!-}\!)}\!(G) Y\!
\right]\!.
\label{ccamplitude}
\end{eqnarray}
Following the same steps as in  \cite{Bianchi:2010zs}, this gives the un-normalized amplitude as
\begin{eqnarray}
W(z)&=& 
\sum_{j} { (2j+1)} \,\frac{N_o}{j^3}\ e^{-2t\hbar j(j+1)- i zj-i\lambda \vol_o j^{\frac32}} 
\label{amplitude}
\end{eqnarray}
where $z=\alpha c+i\beta p$, being $\alpha$ and $\beta$ constants that can be determinated. The resulting amplitude can be studied with different techniques, all giving the same result. 

To begin with, we analyze $W(z)$ directly.  Notice first of all that the sum over $j$ is a gaussian sum, which is peaked on the maximum of the real value of the exponent. This is
\be
j\sim j_o=\frac{Im(z)}{4t\hbar}.
\label{jp}
\ee
Then, notice that $W(z)$ is periodic in $c$. In the sum, we expect the oscillating phases to suppress the sum, except when the imaginary part of the exponent vanishes.  This happens for 
\be
 Re(z) + \lambda \vol_o j^{\frac12}=0.
 \label{zero}
\ee
but since we are near the maxima, we can use \eqref{jp} to obtain
\be
  Re(z) =- \lambda \vol_o j_o^{\frac12} = - \lambda \frac{\vol_o}{\sqrt{4t\hbar}}  Im(z)^{\frac12}.  
\ee 
Squaring the above equation we obtain the Friedmann equation 
\be
  \left(\frac{\dot{a}}{a}\right)^2=\frac{\Lambda}{3}, 
        \label{ds}
\ee
where $\Lambda\sim\lambda^2$. The proportionality constant can be computed using the explicit value of the volume of a tetrahedron and the values given in \cite{Bianchi:2010zs}, and reads 
\be
         \Lambda= \left( \frac{2 \lambda\  G\hbar}{^3\!\sqrt3 \ ^3\!\sqrt2} \right)^2 \!\!, 
\ee
that has correctly the dimensions of an inverse area. 
Equation \eqref{ds} is the Friedmann equation in the presence of a cosmological constant $\Lambda$, which is solved by \mbox{de Sitter} spacetime.  We discuss later on the reason for the quadratic relation between $\Lambda$ and $\lambda$.

\begin{figure}[h]
\includegraphics[scale=0.5]{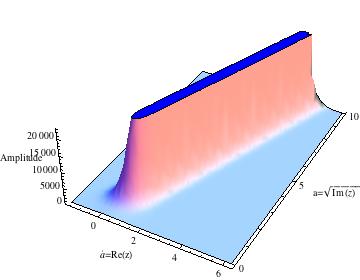}
\caption{Numerical analysis of the transition amplitude \eqref{ccamplitude}. The only free parameter is $\lambda v_o$, here is set equal to $.3$. The computation has been done truncating the sum over $j$ up to a maximum value $j_{max}=200$. This choice is compatible to maximal scale factor plotted.}
\label{prova}
\end{figure}

The validity of the drastic approximations made in this analysis can be confirmed by a numerical analysis of the amplitude \eqref{amplitude}.  For this, we need first to normalize the wave function. The normalization of the holomorphic coherent states has been computed in \cite{Bianchi:2009ky}, and reads (there is a missing $1/2$ factor in \cite{Bianchi:2009ky})
\be
N(p):=
(2\pi t)^{-\frac32}e^{-\frac t2}
\ e^{-\frac{(p/2)^2}{t}} \frac{\sinh
  \left(\frac{p}{2}\right)}{p/2}
   \ee
Plotting the modulus square on the normalized amplitude in the complex $z$ plane
\be
A(z)=|W(z)/N(z)|, 
\ee
we see that it is sharply peaked on a line. The plot of this amplitude in the coordinates $Re(z)\sim   \gamma  \dot  a$ and $\sqrt{Im(z)}\sim a$ is given in \fref{prova}: it clearly shows the linear relation between  $ \dot a$
and $a$ which is characteristic of de Sitter cosmology. 

Furthermore, let us also derive the classical limit using the technique that was used in \cite{Bianchi:2010zs}, to confirm the overall coherence.  The sum over $j$ is still peaked on a value $j_0$. Expanding around  $j_0$  gives 
\be
     i\lambda \vol_o j^{\frac32}\sim  i\lambda \vol_o j_o^{\frac32}+
     \frac32 i\lambda \vol_o j_o^{\frac12}\delta j. 
\ee
The first term is an irrelevant phase. The second amounts to a shift 
\be
 z\to z+ \frac32  \lambda \vol_o j_o^{\frac12}. 
\ee
We therefore recover the result of  \cite{Bianchi:2010zs} up to this shift. In particular, the amplitude derived in   \cite{Bianchi:2010zs} satisfies a Hamiltonian constraint equation that in the classical limit reduces to 
\be
    z^2+\overline z^2=0.
\ee
With the cosmological constant, this becomes
\be
    ( z+ \frac32  \lambda \vol_o j_o^{\frac12})^2+\overline{( z+ \frac32  \lambda \vol_o j_o^{\frac12})^2}=0.
\ee
Using $z=\alpha c+i\beta p\sim \alpha \gamma\dot a+i \beta a^2$, this gives
\be
i4\, a^2\ (\alpha \gamma \dot a + \frac32 \lambda \vol_o j_o^{\frac12} )=0,
\ee
but in our model the volume of the universe (that is a 3-sphere) is twice the volume of the regular tetrahedron namely $a^3\sim 2\vol_o  j_o^{\frac32}$. Therefore we obtain again the Friedmann equation for the de Sitter universe \eqref{ds}.

There are two questions that we must address before concluding. The first is the reason of the surprising quadratic dependence of $\Lambda$ from $\lambda$: if the term added in \eqref{modificazione} is just the cosmological term in the hamiltonian, why is it multiplied by $\sqrt{\Lambda}$ instead of $\Lambda$?  The answer is that the term that must appear in the amplitude is not really $\Lambda V_3\sim \Lambda a^3$, but rather the value of the Hamilton function over the interval covered by the vertex. Let us compute the Hamilton function of the cosmological theory with cosmological constant. 
This can be done evaluating the Einstein-Hilbert action with Hawking boundary term,
\be
S=\int (R-2\Lambda)\sqrt{g}\, + \int_{t_f-t_i} K\sqrt{h}\;,
\ee
on a homogeneous and isotropic metric with scale factor $a(t)$. When the scale factor satisfies the Friedmann equation (\ref{ds}), the bulk term in the action vanishes and the boundary term reduces to
\be
S=\frac23\sqrt{\frac\Lambda3}( a_f^3-a_i^3)\;,
\ee
where $a_i$ and $a_f$ are the initial and final scale factors. Therefore the contribution of the cosmological term to the Hamilton function, and hence to the vertex, is proportional to  $\sqrt{\Lambda} a^3$, and not to $\Lambda a^3$. This modification also accounts for the correct change of dimension from the four-volume term in the action. 

The second question regards the sign of $\lambda$-term. It is clear from the above that this sign must be determined by the kind of solution being considered. Expanding or contracting solutions have different signs.  An amplitude that contains both possibilities can be obtained simply by summing the two terms, in the same manner in which the quantum gravity vertex without cosmological constant is the sum of two terms with opposite time orientation. Indeed, if we replace the exponential of the $\lambda$-term in \eqref{modificazione} with a cosine, we obtain a time symmetric amplitude, whose modulus is given by Figure 2, we shows the two branches of the solution of the Friedmann equation. 

\begin{figure}[h]
\includegraphics[scale=0.8]{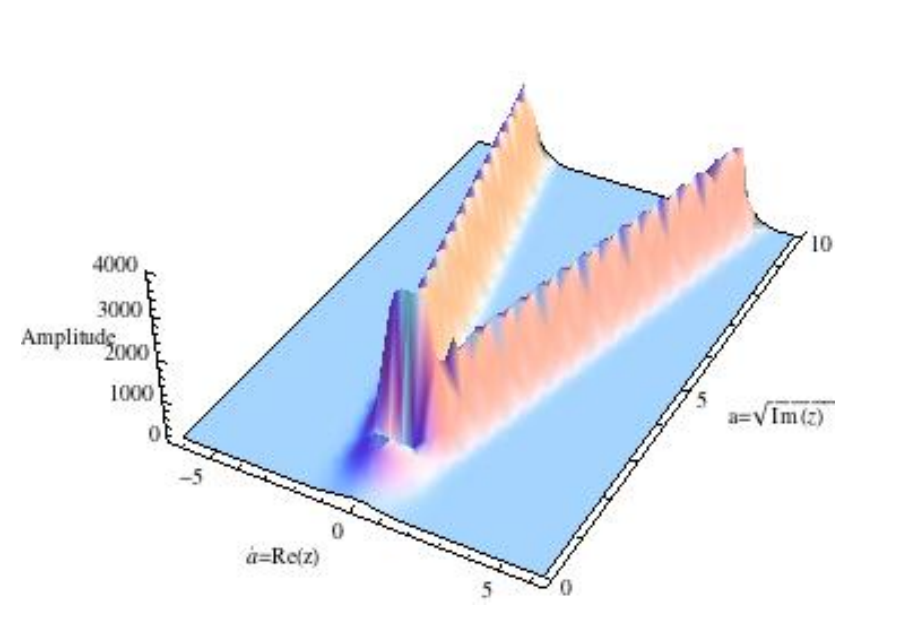}
\caption{Numerical analysis of the transition amplitude with the cosine replacing the exponential.}
\label{prova2}
\end{figure}

\section{Conclusions and summary}

In summary, we have obtained two results. First, we have given a definition of a simple modification of the loop gravity vertex amplitude, which adds a cosmological constant to the theory. Second, we have applied this modified amplitude to cosmology and shown that it leads to the de Sitter spacetime in the large distance limit, where the effect of the curvature can be neglected. 

This is a new element of evidence supporting the conjecture that the loop-gravity amplitude does indeed define a theory whose classical limit is general relativity. It is a stronger result than the analogous result in  \cite{Bianchi:2010zs} because there only flat space-time was recovered, while here a non-trivial solution of the Einstein equations is derived from full quantum gravity.

A major advantage with respect to the analogous calculation in  \cite{Bianchi:2010zs} is that the large distance approximation is now fully justified, as the regime where the curvature term $k/a^2$ is negligible with respect to the cosmological term $\Lambda/3$ -- as in the universe in which we live.

We leave several problems open. First, to study the precise relation between \eqref{modificazione} and  the $q$-deformed vertex amplitude defined in \cite{Fairbairn:2010cp,Han:2010pz}. Second, to include the intrinsic curvature term, which is present because we have assumed a compact slicing; this can be done following \cite{Magliaro:2010qz}, where a pentagonal triangulation of space is used, more appropriate than the dipole one introduced in \cite{Rovelli:2008dx}. Third, to connect this formalism to other  analyses of the cosmological constant in loop quantum cosmology, such as \cite{Bojowald:2008by,Bentivegna:2008bg,Kaminski:2009pc,Bojowald:2010qm}.
Finally, to relate the way the cosmological constant appears here with the way it is has been argued to appear in the spinfoam expansion of canonical loop cosmology \cite{Ashtekar:2009dn,Rovelli:2009tp,Ashtekar:2010ve,Campiglia:2010jw,Henderson:2010qd}.


\subparagraph{Acknolegements}
The work of E.B. is supported by a Marie Curie Intra-European Fellowship within the 7th European Community Framework Programme.
\vfill


\end{document}